\documentclass[sigconf,nonacm,review=false]{acmart}

\usepackage{listings}
\usepackage{longtable}
\usepackage{booktabs}
\usepackage{colortbl}
\usepackage{makecell}
\usepackage{siunitx}
\definecolor{winnerviolet}{rgb}{0.94, 0.90, 0.97}
\colorlet{winnercell}{winnerviolet}  
\usepackage{tikz}
\usetikzlibrary{arrows.meta,positioning,calc,shapes.geometric,backgrounds,fit}
\usepackage[listings,most]{tcolorbox}

\let\origunderscore\_
\renewcommand{\_}{\origunderscore\linebreak[1]}

\lstdefinelanguage{json}{
  morestring=[b]",
  stringstyle=\color{violet!60!black},
  morecomment=[l]{//},
  commentstyle=\color{black!50}\itshape,
  literate=
    *{:}{{{\color{black!70}{:}}}}{1}
    {,}{{{\color{black!50}{,}}}}{1}
    {\{}{{{\color{violet!70!black}{\{}}}}{1}
    {\}}{{{\color{violet!70!black}{\}}}}}{1}
    {[}{{{\color{violet!70!black}{[}}}}{1}
    {]}{{{\color{violet!70!black}{]}}}}{1},
}

\newtcblisting{jsonbox}[1]{
  listing only,
  listing options={language=json, basicstyle=\ttfamily\scriptsize,
    breaklines=true, breakatwhitespace=false, breakindent=10pt,
    postbreak=\mbox{\hspace{0pt}$\hookrightarrow$\space},
    columns=fullflexible, keepspaces=true, showstringspaces=false},
  colback=violet!4,
  colframe=violet!60!black,
  coltitle=white,
  colbacktitle=violet!60!black,
  fonttitle=\sffamily\bfseries\small,
  title={#1},
  boxrule=0.5pt,
  arc=2pt,
  left=4pt, right=4pt, top=2pt, bottom=2pt,
  width=\linewidth,
  before=\par\medskip\noindent,
  after=\par\medskip,
  enhanced, breakable,
}

\newtcolorbox{defbox}[2][]{
  colback=white,
  colframe=black!70,
  coltitle=black,
  colbacktitle=white,
  fonttitle=\sffamily\bfseries\small,
  title={#2},
  boxrule=0.6pt,
  arc=0pt,
  left=6pt, right=6pt, top=4pt, bottom=4pt,
  width=\linewidth,
  before=\par\medskip\noindent,
  after=\par\medskip,
  enhanced, breakable,
  attach boxed title to top left={xshift=8pt, yshift=-\tcboxedtitleheight/2},
  boxed title style={colback=white, colframe=white, boxrule=0pt,
                     left=2pt, right=2pt, top=0pt, bottom=0pt},
  #1
}

\newtcblisting{pybox}[1]{
  listing only,
  listing options={language=Python, basicstyle=\ttfamily\scriptsize,
    breaklines=true, breakatwhitespace=false, breakindent=10pt,
    postbreak=\mbox{\hspace{0pt}$\hookrightarrow$\space},
    columns=fullflexible, keepspaces=true, showstringspaces=false,
    keywordstyle=\color{teal!70!black}\bfseries,
    stringstyle=\color{violet!60!black},
    commentstyle=\color{black!50}\itshape,
  },
  colback=teal!4,
  colframe=teal!60!black,
  coltitle=white,
  colbacktitle=teal!60!black,
  fonttitle=\sffamily\bfseries\small,
  title={#1},
  boxrule=0.5pt,
  arc=2pt,
  left=4pt, right=4pt, top=2pt, bottom=2pt,
  width=\linewidth,
  before=\par\medskip\noindent,
  after=\par\medskip,
  enhanced, breakable,
}

\providecommand{\tightlist}{\setlength{\itemsep}{0pt}\setlength{\parskip}{0pt}}

\setcopyright{none}
\acmConference{}{}{}
\acmISBN{}
\acmDOI{}
\acmYear{}
\copyrightyear{}
\renewcommand\footnotetextcopyrightpermission[1]{}
\title{Self-Reflective APIs: Structure Beats Verbosity\\ for AI Agent Recovery}

\author{Arquimedes Canedo}
\orcid{0000-0003-3506-6563}
\affiliation{\institution{Siemens Digital Industries Software, USA}\country{}}

\author{Grama Chethan}
\affiliation{\institution{Siemens Digital Industries Software, USA}\country{}}

\begin{document}

\begin{abstract}
When an AI agent calls an API and hits a validation error, it needs more than \emph{what} went wrong---it needs \emph{what to do next}. A \emph{self-reflective API} returns, on validation failure, a machine-readable \texttt{recovery\_feedback.suggestions[]} payload sufficient for the agent to repair the request and retry without external reasoning. On a leak-audited pilot ($N{=}30$ per cell, 3 LLMs, 10 adversarial tasks), structured suggestions lift task-completion rate by $+36.7$--$40.0$pp over plain-English diagnoses on Anthropic models (Fisher's exact $p \le 0.0022$), at $1.8$--$2.2\times$ better per-success token efficiency. The lift is not significant on gpt-4o-mini ($p{=}0.435$); a second-domain replication on a billing API confirms the pattern. The comparison only holds after auditing two undocumented classes of answer leakage in LLM benchmarks. We ship \texttt{audit\_prompt\_leakage.py} as reusable CI infrastructure. Code and data: \url{https://github.com/arquicanedo/self-reflective-apis}.

\end{abstract}


\maketitle

\section{Introduction and Motivation}\label{introduction-and-motivation}

LLMs now call APIs as autonomous agents, and they fail differently than humans do. A developer reads an error, infers the missing context, and edits the request. An agent reads the same error and burns tokens guessing until it gives up. Today's APIs reply with generic codes and short prose written for humans, leaving the agent to invent the fix from training-data priors. When the constraint is proprietary (a company-specific rule, a certification requirement, a cultural standard, real-time state) those priors do not contain the answer.

This is not a functional gap. The API knows the rule and just does not say what to do about it, because generic error codes and free-text messages cannot carry the structured, actionable guidance an agent needs to self-correct. The cost lands hardest where APIs encode proprietary business logic, the kind of rules that exist only inside an organization and that a general-purpose LLM has never seen. Modern LLMs can guess solutions to generic problems, but they cannot guess proprietary constraints.

We call the fix \textbf{self-reflective APIs}:

\begin{defbox}{Definition: Self-Reflective API}\label{def:self-reflective}
A \textbf{self-reflective API} is one whose error responses contain a machine-readable repair payload---structured suggestions naming the specific parameter changes required---sufficient for an autonomous agent to recover from a domain-specific validation rejection without human intervention or external reasoning.
\end{defbox}

\noindent This contract is narrower than ``rich error reporting'' (RFC 7807, JSON:API) and narrower than agent-side reflection (Reflexion, ReAct). A self-reflective API moves the repair hint from prose into a typed payload the next LLM call can consume directly. The payoff lands where APIs hold knowledge the LLM does not, in proprietary domains.

\subsection{The LLM Intelligence Challenge and Our Research Focus}\label{the-llm-intelligence-challenge-and-our-research-focus}

One observation shaped the evaluation. Modern LLMs already know how to fix many simple validation errors without being told. An unaudited probe on generic adversarial tasks like ``convert to gluten-free without providing gluten-free flour'' had Traditional and Reflective modes performing near-identically on a GPT-4-class model. We do not report this probe as a result (it predates the leak audit in Section~\ref{leakage-audit}, and the validator's own error string may have contained the answer), but it pointed us at the right question. Self-reflective feedback only matters when the LLM cannot guess the fix.

That narrows the thesis. Self-reflective APIs pay off when validation depends on \emph{domain knowledge LLMs cannot reasonably infer from training data}. The targets in this paper are context-dependent cultural rules (coconut milk is incompatible with French cuisine), proprietary certifications (\emph{Bob's Red Mill Certified Gluten Free 1-to-1 Baking Flour}, not generic ``gluten-free flour''), specialized technical knowledge (sourdough requires a wild-yeast starter), numerical-precision requirements (scaling by $1.625\times$ yields 3.25 cups), and cascading validation dependencies where fixing one error reveals the next.

Our experiments cover only context-dependent scenarios. We compare three LLMs from two vendors (claude-haiku-4-5, claude-sonnet-4-6, gpt-4o-mini) against three error-detail levels, separating the contribution of \emph{structured machine-readable suggestions} from \emph{plain-English error messages}. This is a pilot, not a definitive evaluation. One domain (recipe conversion), three models, ten adversarial tasks, one retry-loop agent (Section~\ref{external-validity} covers threats to validity). Build self-reflection where APIs hold proprietary business logic, external state, and specialized rules. Skip it for generic validation LLMs already understand.

\section{Related Work}\label{related-work}

GraphQL's introspection \cite{ref1} is the closest precedent. Clients query the schema to learn what operations exist. But GraphQL introspection is structural. It says what is possible, not why a call failed or how to fix it. Self-reflective APIs add behavioral reflection. The API names the parameter changes that would make the next call succeed.

JSON:API \cite{ref2} and RFC 7807 Problem Details \cite{ref3} standardize rich error reporting, but they describe what broke, not what to do next. Observability and structured-logging work \cite{ref4} targets human operators, not agents. The intersection of API design and AI has mostly looked at discovery and integration \cite{ref5}, not the recovery loop.

\paragraph{Relation to LLM tool use and structured outputs.} One could argue self-reflective APIs are already covered by function calling with JSON-Schema \cite{ref16-funcall}, structured-output validation, and tool ecosystems like Gorilla \cite{ref17-gorilla} and Toolformer \cite{ref18-toolformer}. Two distinctions matter. JSON-Schema describes \emph{static} shape and per-field constraints. It cannot express the cross-field, context-dependent rules this work targets (``coconut milk is incompatible with French cuisine,'' ``oats also need certification once flour is fixed''). And tool-use benchmarks measure whether the model picks the right tool with the right shape, not whether an agent can \emph{recover} from a semantic rejection at the application layer. Self-reflective APIs target the recovery loop after the request is well-formed but rejected.

\paragraph{Reasoning and reflection in agents.} Chain-of-Thought \cite{ref9}, ReAct \cite{ref10}, and Reflexion \cite{ref11} improve how the agent thinks. We change what the API tells the agent. Agent-side reflection still operates on whatever information the API chose to expose. Combining the two is open.

\section{Design: Self-Reflective API Framework}\label{design-self-reflective-api-framework}

Self-reflective APIs treat structured semantic feedback as part of the API contract, not an afterthought. The framework defines two reflection types we evaluate (recovery guidance as primary, and intent disambiguation) and proposes a third (confidence signaling) we do not yet test. Recovery guidance names the parameter changes. Intent disambiguation handles ambiguous or incomplete requests. Feedback travels with the response payload, not in a separate channel.

The core idea is simple. An API that knows why it rejected a request should say so in a form the next LLM call can act on. Structured feedback objects sit alongside the regular payload and carry machine-readable guidance for retry, parameter changes, or endpoint selection. The overhead is small enough not to compromise high-throughput use.

\subsection{Framework Architecture}\label{framework-architecture}

\subsubsection{Self-Reflective API Schema v0.1}\label{schema-v0-1}

We name the wire-format contract \textbf{Self-Reflective API Schema v0.1} so future work can cite, extend, or implement it as a discrete artifact. A response is conformant if, on validation failure, it carries a top-level \texttt{recovery\_feedback} object with (i)~a \texttt{type} discriminator, (ii)~a per-rule \texttt{message} field, and (iii)~a \texttt{suggestions[]} array whose entries each name an action \texttt{type} from a registered vocabulary and a \texttt{parameters} object holding the literal values an agent should merge into its next request. The schema below is the normative form.

\subsubsection{Core Response Structure}\label{core-response-structure}

The envelope is deliberately small. A \texttt{data} field for the original payload, a \texttt{feedback} object that surfaces only when the API has something to say, and a standard \texttt{metadata} block. The \texttt{feedback.type} discriminator tells the agent which of the three reflection modes to dispatch on. \texttt{suggestions[]} is the load-bearing field, an ordered list of concrete actions the agent can apply without re-prompting an LLM.

\begin{jsonbox}{Schema v0.1 -- Core response structure}
{
  "data": "...",                    // Original payload
  "feedback": {                     // Semantic feedback object
    "type": "INTENT_DISAMBIGUATION | RECOVERY_GUIDANCE | CONFIDENCE_SIGNAL",
    "message": "Human-readable explanation",
    "structured_data": {...},       // Machine-readable context
    "confidence": 0.95,             // 0.0-1.0 confidence score
    "suggestions": [                // Recommended actions
      {
        "type": "RETRY | MODIFY_PARAMS | TRY_DIFFERENT_ENDPOINT",
        "description": "What to do",
        "parameters": {...}         // How to do it
      }
    ]
  },
  "metadata": {...}                 // Standard headers, timing
}
\end{jsonbox}

\subsubsection{Types of Self-Reflection}\label{types-of-self-reflection}

Our framework provides two types of structured semantic feedback. \emph{Intent disambiguation} identifies ambiguous or incomplete requests and offers structured alternatives so the agent can refine its parameters before retrying. \emph{Recovery guidance} (this paper's primary focus) returns actionable steps for failed operations. A root-cause diagnosis paired with specific domain-aware fixes drawn from a fixed vocabulary of machine-readable action types (\texttt{ADD\_INGREDIENT}, \texttt{CLARIFY\_MEASUREMENT}, \texttt{REPLACE\_INCOMPATIBLE\_INGREDIENT}, \texttt{MODIFY\_PARAMS}, \texttt{USE\_SPECIFIC\_BRAND}, \texttt{FIX\_SCALING\_PRECISION}) that an automated retry loop can execute without human intervention.

\subsection{Adopting Self-Reflective APIs in \texorpdfstring{$\sim$}{\textasciitilde}20 Lines}\label{minimal-adoption}

The minimum viable self-reflective endpoint is small. The snippet below turns a FastAPI handler into a Schema v0.1 conformant one by wrapping its validation step. The same pattern works for any framework that returns JSON.

\begin{pybox}{Minimal self-reflective endpoint (FastAPI, $\sim$20 lines)}
from fastapi import FastAPI, Response
app = FastAPI()

def validate(req):  # -> list of (err_type, msg, fix_action, fix_params)
    ...

@app.post("/recipe/convert")
def convert(req: dict, response: Response):
    failures = validate(req)
    if not failures:
        return {"data": do_convert(req)}
    response.status_code = 422
    return {
      "success": False,
      "validation_errors": [
        {"type": t, "message": m} for t, m, _, _ in failures
      ],
      "recovery_feedback": {
        "type": "RECOVERY_GUIDANCE",
        "message": "; ".join(m for _, m, _, _ in failures),
        "suggestions": [
          {"type": act, "description": m, "parameters": p}
          for _, m, act, p in failures
        ],
      },
    }
\end{pybox}

\noindent Two adopter requirements: (1)~each validation rule emits a typed action plus the literal parameters needed to repair the request, and (2)~the action \texttt{type} comes from a vocabulary the agent has been told about (via the OpenAPI spec or a one-shot system prompt). Everything else is optional.

\section{Implementation}\label{implementation}

\subsection{System Architecture}\label{system-architecture}

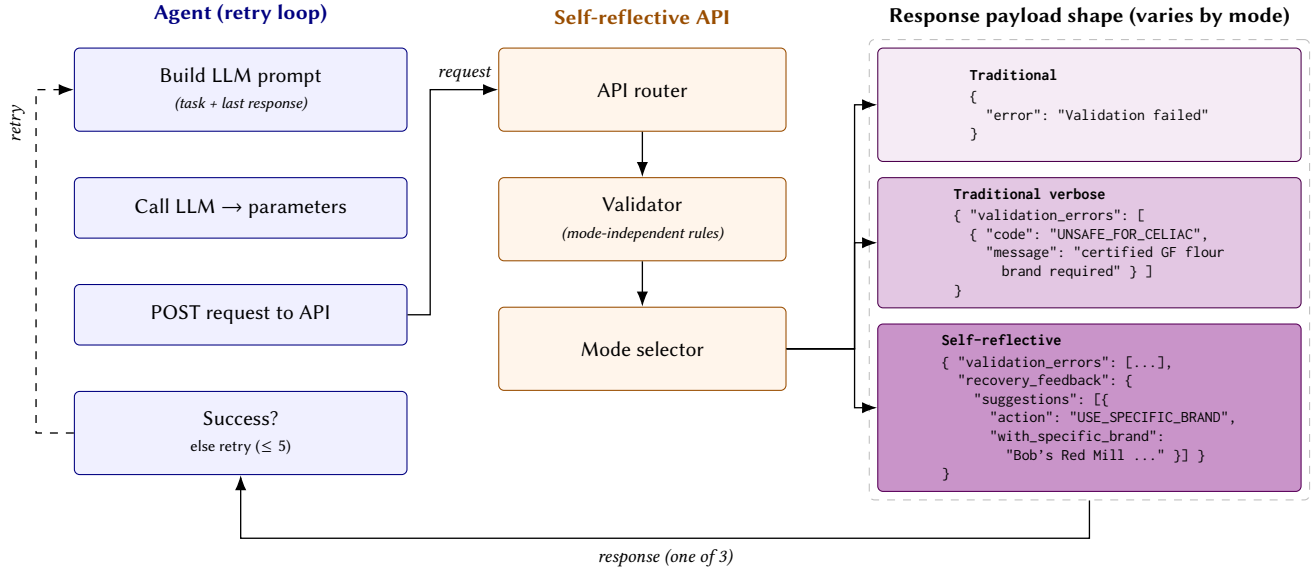
\begin{figure*}[t]
\centering
\begin{tikzpicture}[
  font=\sffamily\small,
  node distance=6mm and 9mm,
  box/.style={draw, rounded corners=2pt, align=center, inner sep=4pt, minimum height=8mm, fill=white},
  agentbox/.style={box, fill=blue!6, draw=blue!50!black},
  apibox/.style={box, fill=orange!8, draw=orange!60!black},
  modebox/.style={draw, rounded corners=2pt, align=left, inner sep=4pt, font=\ttfamily\scriptsize, minimum width=46mm, minimum height=15mm, anchor=north west},
  trad/.style={modebox, fill=violet!8,  draw=violet!60!black},
  verb/.style={modebox, fill=violet!22, draw=violet!60!black},
  refl/.style={modebox, fill=violet!42, draw=violet!70!black},
  arr/.style={-{Latex[length=2mm]}, line width=0.5pt},
  dim/.style={font=\footnotesize\itshape, text=black},
]

\node[agentbox, minimum width=44mm, minimum height=11mm]
  (build) {Build LLM prompt\\\scriptsize\itshape (task + last response)};
\node[agentbox, below=of build, minimum width=44mm, minimum height=8mm]
  (call) {Call LLM \(\to\) parameters};
\node[agentbox, below=of call, minimum width=44mm, minimum height=8mm]
  (post) {POST request to API};
\node[agentbox, below=of post, minimum width=44mm, minimum height=11mm]
  (decide) {Success?\\\scriptsize else retry ($\le 5$)};

\draw[arr, dashed] (decide.west) -- ++(-5mm,0)
  |- node[dim, pos=0.45, left=1pt, rotate=90, anchor=south] {retry}
  (build.west);

\node[apibox, minimum width=38mm, minimum height=11mm,
      anchor=north west] at ([xshift=12mm]build.north east)
  (router) {API router};
\node[apibox, below=of router, minimum width=38mm, minimum height=11mm,
      align=center]
  (validator) {Validator\\\scriptsize\itshape (mode-independent rules)};
\node[apibox, below=of validator, minimum width=38mm, minimum height=11mm,
      align=center]
  (sel) {Mode selector};

\draw[arr] (post.east) -- ++(3mm,0) |- node[dim, above, pos=0.75] {request} (router.west);
\draw[arr] (router.south) -- (validator.north);
\draw[arr] (validator.south) -- (sel.north);

\node[trad, anchor=north west, minimum width=56mm]
  at ([xshift=12mm]router.north east)
  (tradbox)
  {\textbf{Traditional}\\[1pt]
   \{\\
   \ \ "error": "Validation failed"\\
   \}};
\node[verb, below=2mm of tradbox.south west, anchor=north west,
      minimum width=56mm]
  (verbbox)
  {\textbf{Traditional verbose}\\[1pt]
   \{ "validation\_errors": [\\
   \ \ \{ "code": "UNSAFE\_FOR\_CELIAC",\\
   \ \ \ \ "message": "certified GF flour\\
   \ \ \ \ \ \ brand required" \} ]\\
   \}};
\node[refl, below=2mm of verbbox.south west, anchor=north west,
      minimum width=56mm]
  (reflbox)
  {\textbf{Self-reflective}\\[1pt]
   \{ "validation\_errors": [...],\\
   \ \ "recovery\_feedback": \{\\
   \ \ \ \ "suggestions": [\{\\
   \ \ \ \ \ \ "action": "USE\_SPECIFIC\_BRAND",\\
   \ \ \ \ \ \ "with\_specific\_brand":\\
   \ \ \ \ \ \ \ \ "Bob's Red Mill ..." \}] \}\\
   \}};

\draw[arr] (sel.east) -| ([xshift=-3mm]tradbox.west) -- (tradbox.west);
\draw[arr] (sel.east) -| ([xshift=-3mm]verbbox.west) -- (verbbox.west);
\draw[arr] (sel.east) -| ([xshift=-3mm]reflbox.west) -- (reflbox.west);

\begin{pgfonlayer}{background}
  \node[draw=black!25, dashed, rounded corners=2pt,
        fit=(tradbox)(verbbox)(reflbox), inner sep=3pt] (respgrp) {};
\end{pgfonlayer}

\draw[arr]
  (respgrp.south) -- ++(0,-5mm)
  -| node[dim, pos=0.25, below=1pt] {response (one of 3)}
  (decide.south);

\node[font=\sffamily\bfseries\small, above=2mm of build, text=blue!50!black]
  {Agent (retry loop)};
\node[font=\sffamily\bfseries\small, above=2mm of router, text=orange!60!black]
  {Self-reflective API};
\node[font=\sffamily\bfseries\small] at ($(tradbox.north)+(0,4mm)$)
  {Response payload shape (varies by mode)};

\end{tikzpicture}
\caption{Experimental control flow. A retry-loop agent ($\le 5$ attempts) calls the API. The validator and business-logic layer are mode-independent. The only thing that varies across the three evaluation conditions is the \emph{shape} of the response payload on validation failure. \textit{Traditional} returns a generic string, \textit{Traditional verbose} adds a structured per-rule diagnosis without naming the fix, and \textit{Self-reflective} additionally returns a machine-readable \texttt{recovery\_feedback.suggestions[]} array carrying the specific parameters the agent should pass on the next call. Examples shown are abbreviated. Full schemas in Section~\ref{response-structure}.}
\label{fig:architecture}
\end{figure*}

Figure~\ref{fig:architecture} shows the experimental control flow. A retry-loop agent ($\le 5$ attempts) calls the API. The validator and business-logic layer are identical across all three modes, so the only thing that varies is the \emph{shape} of the response payload returned on failure. Traditional returns a generic error string. Verbose returns the same per-rule diagnoses without literal fix values. Reflective returns those diagnoses plus a machine-readable \texttt{recovery\_feedback.suggestions[]} array with the specific parameters needed to retry. The agent merges any returned suggestion parameters into its next request and re-invokes the LLM with the updated context. Mode is a pure response-formatting toggle. It does not change which validation rules fire or which inputs are accepted.

\subsection{Reference Implementation: Recipe API}\label{reference-implementation-recipe-api}

We built a Recipe API for recipe conversion and ingredient substitution as the testbed. The domain has rich validation, domain-specific constraints, and natural openings for proprietary business logic that general-purpose LLMs cannot infer. Both traditional and reflective modes share identical underlying functionality so the comparison stays fair.

\emph{Design philosophy: context-dependent validation.} We chose validation rules because they require knowledge LLMs lack. Cultural authenticity (French cuisine forbids coconut milk), medical-certification brands (celiac-safe options like Bob's Red Mill Certified GF), specialized technique (sourdough requires a wild-yeast starter), numerical precision (exact scaling ratios like $1.625\times$), and cascading dependencies (fixing flour reveals oats also need certification). This keeps the experiments on scenarios where feedback adds measurable value, not generic validation LLMs already handle.

\subsubsection{API Endpoints and Validation Rules}\label{api-endpoints}\label{domain-specific-validation-rules}

The Recipe API exposes two endpoints, \texttt{POST /api/recipe/convert} (recipe conversion under dietary, cuisine, measurement, or scaling constraints) and \texttt{POST /api/recipe/substitute} (ingredient replacement), each supporting both traditional and reflective response modes via a \texttt{mode} query parameter. Full request/response schemas are in the OpenAPI spec shipped with the repo.

The validator encodes five families of domain-specific rules chosen because they require knowledge a general-purpose LLM cannot reliably infer. (1)~\emph{Context-dependent ingredient compatibility} (e.g.\ coconut milk incompatible with French or Italian cuisine, instant yeast incompatible with sourdough). (2)~\emph{Celiac-safe certification requirements} naming specific brands (e.g.\ \emph{Bob's Red Mill Certified Gluten Free 1-to-1 Baking Flour}) rather than generic ``gluten-free''. (3)~\emph{Numeric precision} for non-standard scaling ratios (e.g.\ $4 \to 6.5$ servings requires \texttt{3.25 cups}, not rounded \texttt{3 cups}). (4)~\emph{Measurement specificity} (rejecting ``a handful of flour''). (5)~\emph{Missing-alternative validation} for dietary conversion requests. The full validator source, including the \texttt{INCOMPATIBLE\_COMBINATIONS} matrix and the \texttt{CELIAC\_SAFE\_BRANDS} table, is in the repo.

\subsubsection{Response Structure}\label{response-structure}

\begin{jsonbox}{Traditional mode -- success response}
{
  "data": {
    "converted_recipe": {
      "ingredients": ["2 cups Bob's Red Mill Certified GF Flour", ...],
      "instructions": [...],
      "servings": 8
    }
  }
}
\end{jsonbox}

\begin{jsonbox}{Traditional mode -- validation failure}
{
  "success": false,
  "message": "Validation failed: Missing gluten-free flour alternative",
  "validation_errors": [
    {
      "type": "MISSING_ALTERNATIVE",
      "ingredient": "all-purpose flour",
      "message": "No gluten-free alternative provided for flour-based ingredient"
    }
  ]
}
\end{jsonbox}

\begin{jsonbox}{Self-reflective mode -- validation failure}
{
  "success": false,
  "message": "Validation failed: Missing gluten-free flour alternative",
  "validation_errors": [
    {
      "type": "MISSING_ALTERNATIVE",
      "ingredient": "all-purpose flour",
      "message": "No gluten-free alternative provided for flour-based ingredient"
    }
  ],
  "recovery_feedback": {
    "analysis": "Recipe contains wheat-based ingredients incompatible with gluten-free dietary restriction",
    "suggestions": [
      {
        "type": "ADD_INGREDIENT",
        "description": "Add gluten-free flour alternative to recipe ingredients list",
        "suggested_ingredient": "gluten-free flour (certified brand recommended for celiac safety)",
        "example_modification": {
          "ingredients": ["2 cups gluten-free flour", "existing ingredients..."]
        }
      }
    ],
    "recovery_likelihood": "high"
  }
}
\end{jsonbox}

The key difference in reflective mode is the \texttt{recovery\_feedback} object containing structured suggestions with machine-readable action types (\texttt{ADD\_INGREDIENT}, \texttt{CLARIFY\_MEASUREMENT}, \texttt{REPLACE\_INCOMPATIBLE\_INGREDIENT}, \texttt{MODIFY\_PARAMS}) and concrete examples of how to modify the request.

All experiments run in \texttt{validation=strict} (every rule enforced). The validator and the feedback generator are separate modules so the reflective payload is an additive layer rather than entangled with business logic.

\subsection{Model Selection and Cost Trade-offs}\label{model-selection-and-cost-trade-offs}

We test three LLMs from two vendors (claude-haiku-4-5, claude-sonnet-4-6, gpt-4o-mini) spanning roughly an order of magnitude in per-token price. Two questions. Does the effect generalize across capability tiers and vendors, and does self-reflective feedback let a cheaper model close part of the gap to a more expensive one?

\subsubsection{Per-Model Token-Efficiency Effect}\label{cost-comparison}

Reflective's success-rate and token-efficiency gains over Verbose, the contribution attributable to \emph{structured machine-readable suggestions} over \emph{plain-English diagnoses}, are large and significant on the Anthropic models ($+36.7$ to $+40.0$pp, $1.8\times$ on haiku, $2.2\times$ on sonnet) and small or tied on gpt-4o-mini ($+13.3$pp, $p=0.435$, per-success tokens 3{,}548 vs 3{,}665). Per-cell figures are in Tables~\ref{tab:success3x3}, \ref{tab:tokens3x3}, and \ref{tab:fisher} below.

\subsubsection{Decomposing the cost reduction}\label{sec:cost-decomposition}

The intrinsic contribution (the per-model token-efficiency factor) is independent of model choice (Table~\ref{tab:cost-decomp}). Swapping an expensive model for a cheaper one (e.g.\ GPT-4 $\to$ gpt-4o-mini, roughly $200\times$ on vendor pricing) compounds a separate factor on top. We do not attribute that to self-reflection.

\begin{table}[h]
\centering
\small
\caption{Decomposition of cost reduction. The framework's intrinsic effect is the per-model token-efficiency factor. Any model-swap factor (e.g.\ $\sim200\times$ for GPT-4 $\to$ gpt-4o-mini) is independent of self-reflection and would apply to any technique sufficient to make a cheaper model succeed.}
\label{tab:cost-decomp}
\renewcommand{\arraystretch}{1.15}
\begin{tabular}{lcl}
\toprule
Component & Factor & Source \\
\midrule
\makecell[l]{Token efficiency, haiku-4-5\\\scriptsize Refl.\ vs Verbose}    & $1.8\times$       & \S\ref{sec:eval-tokens} \\
\makecell[l]{Token efficiency, sonnet-4-6\\\scriptsize Refl.\ vs Verbose}   & $2.2\times$       & \S\ref{sec:eval-tokens} \\
\makecell[l]{Token efficiency, gpt-4o-mini\\\scriptsize Refl.\ vs Verbose}  & $1.0\times$ tied & \S\ref{sec:eval-tokens} \\
\midrule
\makecell[l]{Per-token model swap\\\scriptsize GPT-4 $\to$ gpt-4o-mini, illustrative} & ${\sim}200\times$ & Vendor pricing \\
\bottomrule
\end{tabular}
\end{table}

Self-reflective APIs contribute the per-model token-efficiency factor and the $+13.3$ to $+40.0$pp success-rate gain over plain-English diagnoses (significant on Anthropic, not on gpt-4o-mini). The model-swap factor is for total-cost-of-ownership planning only. It is not part of the framework effect.

\section{Evaluation}\label{evaluation}

\subsection{Experimental Design}\label{experimental-design}

\subsubsection{Hypothesis}\label{hypothesis-1}

\textbf{Primary:} AI agents using self-reflective APIs will achieve higher task-completion rates than agents using either generic or plain-English-verbose error responses, on tasks requiring context-dependent validation knowledge.

\textbf{Secondary:} Self-reflective APIs will reduce mean retry count and per-success token cost compared to either baseline.

We measure all three (success rate, retry count, tokens-per-success). Success rate is the headline because it is what a downstream agent system actually sees. Retry count and token cost are there to show the gains are not bought by simply ``trying more times.''

\subsubsection{Test Conditions}\label{test-conditions}

We compare three error-detail levels, all sharing identical validation logic and identical task inputs.

\emph{Traditional} (floor) returns a single generic error message (``recipe validation failed'') with no per-rule diagnosis. \emph{Traditional Verbose} (intermediate) returns per-rule plain-English diagnoses (e.g., ``coconut milk is not traditional in French cuisine''), with literal fix recipes audited \emph{out} of the message field (Section~\ref{leakage-audit}). \emph{Reflective} (treatment) returns the verbose diagnosis plus a structured \texttt{recovery\_feedback.suggestions[]} payload with machine-readable parameters (\texttt{REPLACE\_INCOMPATIBLE\_INGREDIENT}, \texttt{USE\_SPECIFIC\_BRAND}, etc.) carrying the literal fix. All three modes face identical strict validation. The only differences are the verbosity of the per-error \texttt{message} field and the presence of the structured suggestions payload.

Verbose is the key methodological lever. It isolates \emph{machine-readable structure} from the trivial effect of merely making error text longer. Verbose-vs-Reflective is the core test of the framework. Verbose-vs-Traditional measures the value of \emph{any} per-rule error message.

\subsubsection{Adversarial Task Design}\label{adversarial-task-design}

We built 10 adversarial tasks that fail validation under normal circumstances, require context-dependent knowledge LLMs lack, and test whether structured feedback can guide recovery.

\emph{Design philosophy.} Tasks only test validation needing domain-specific knowledge LLMs cannot infer from training data, not generic validation they can guess. The shift came from finding that sophisticated LLMs hit 93\% success in both Traditional and Reflective on simple tasks where inference is possible.

\emph{Task categories.} The suite spans four families. \emph{Context-dependent ingredient compatibility} (4 tasks) covers French-cuisine authenticity (\texttt{adv\_ctx\_002}, \texttt{adv\_ctx\_004}, coconut milk $\rightarrow$ crème fraîche), Italian-cuisine authenticity (\texttt{adv\_ctx\_003}, vegan cheese $\rightarrow$ nutritional yeast), sourdough leavening (\texttt{adv\_ctx\_006}, instant yeast $\rightarrow$ wild-yeast starter), and meringue preparation (\texttt{adv\_ctx\_005}, baking powder incompatible with egg whites), validating cultural cuisine standards LLMs lack training data for. \emph{Celiac certification requirements} (2 tasks) cover brand-specific certification (\texttt{adv\_ctx\_001}, generic gluten-free flour $\rightarrow$ Bob's Red Mill Certified GF) and cascading validation (\texttt{adv\_cascade\_001}, flour fixed $\rightarrow$ oats also need certification), targeting the gap between knowing gluten-free concepts and knowing specific medically certified brands. \emph{Numerical precision} (3 tasks, \texttt{adv\_scale\_001/002/003}, scaling by $1.625\times$, $1.583\times$, $1.375\times$) requires exact decimal amounts (3.25 cups, not 3 or 3.5) and tests whether smaller models like gpt-4o-mini can apply precise arithmetic from feedback. \emph{Combined multi-validation} (1 task, \texttt{adv\_combo\_001}, French + celiac-safe + vague measurements, difficulty 10.0) tests simultaneous recovery from multiple context-dependent errors.

\subsubsection{Implementation Details}\label{implementation-details}

\emph{Model selection.} Three LLMs from two vendors. \texttt{claude-haiku-4-5} and \texttt{claude-sonnet-4-6} (Anthropic Messages API), and \texttt{gpt-4o-mini} (OpenAI via the LangChain wrapper). Token usage comes from each provider's billed \texttt{usage} field, not character-count estimates. Three models is enough to rule out single-model artifacts and to surface vendor-level differences across baselines.

\emph{Agent framework.} A simple retry-loop agent that parses structured feedback when provided, adjusts request parameters based on recovery guidance, and caps at five retries per task. Deep parameter copying isolates each adjustment so we can attribute success to the specific change. The architecture deliberately mirrors a typical production agent rather than a sophisticated planner or reasoner.

\emph{Validation layer.} Both modes share an identical \texttt{RecipeValidator} with comprehensive rule checking. The API returns HTTP 200 with a \texttt{\{"success": false\}} body for validation failures. Reflective mode appends a structured \texttt{recovery\_feedback} payload to the same error response.

\emph{Metrics collection.} We record logical success (not HTTP status), per-provider billed token usage (Anthropic Messages API, OpenAI via the LangChain wrapper), retry counts, and the recovery actions and parameter adjustments the agent took on each retry. Every metric is captured identically across modes for fair comparison.

\subsection{Benchmark Hygiene: Answer-Leakage Audit}\label{leakage-audit}

One methodology step matters beyond this paper. A systematic audit for \emph{answer leakage} in LLM evaluation benchmarks. An earlier version of this experiment had traditional\_verbose roughly tying reflective, a result that would have inverted our main conclusion. Tracing it exposed two distinct classes of leakage that are easy to introduce, hard to notice, and likely affect other LLM agent benchmarks.

\paragraph{Class 1: validator-message leaks} The validator's plain-English \texttt{message} field originally carried the literal fix. Brand names, precomputed target amounts, replacement-ingredient strings. Verbose and Reflective both delivered the answer, just wrapped differently. The API-layer analogue of test-set contamination. The ``treatment'' field looked strong only because the ``control'' field was already leaking the answer. After the redesign, \texttt{validation\_errors[].message} carries only diagnosis (``a specific certified gluten-free flour brand is required'', ``coconut milk is not traditional in French cuisine'', ``ratio $4 \to 7$ requires exact precision''). The literal prescription (\texttt{Bob's Red Mill Certified Gluten-Free Oats}, \texttt{regular milk or cream}, \texttt{expected\_amount: 3.5}) lives \emph{only} in \texttt{recovery\_feedback.suggestions[]}, populated only in reflective mode.

\paragraph{Class 2: task-prompt leaks} The driver script ships \texttt{task.description} and \texttt{task.success\_criteria} verbatim into the LLM prompt. Author-facing copy meant to remind the experimenter what ``success'' looks like (e.g., \texttt{success\_criteria} reading ``Agent should replace coconut milk with regular milk or cream'') becomes an answer key that fires in \emph{every} mode, inflating Traditional and Verbose alike and shrinking the Reflective lift. We missed it at first because these fields are author-facing in the task library. The audit only caught it after the validator-message fix failed to widen the gap as expected. The class generalizes. Any benchmark where task metadata and the ground-truth validator come from the same author is at risk, because author intent about what ``correct'' means mirrors the validator's acceptance criteria.

\emph{A general taxonomy.} Both classes generalize beyond self-reflective APIs. Any LLM benchmark that (a) compares conditions on how much information reaches the model and (b) is built by the same team that defines correctness must audit both the \emph{response channel} (does the baseline response already carry the answer?) and the \emph{task channel} (does the task prompt already carry the answer?). Skip either, and the experiment measures prompt engineering, not the treatment.

\paragraph{Audit infrastructure} We ship \texttt{experiments/audit\_prompt\_leakage.py} as reusable CI infrastructure. It enumerates known fix values from the validator's class data (\texttt{INCOMPATIBLE\_COMBINATIONS}, \texttt{CELIAC\_SAFE\_BRANDS}) plus a curated substring list catching paraphrases a capable LLM would still recognize (e.g., the literal ratio ``1.625'' for \texttt{adv\_scale\_001}, or the worked example ``11 eggs'' for \texttt{adv\_scale\_003}), then scans every LLM-visible task field. Non-zero exit code on any leak. Run it in CI before any evaluation sweep. Adapt it for your benchmark by enumerating fix values from validator source and scanning all LLM-visible fields. The headline numbers here come from runs after \emph{both} leak classes were fixed. The verbose baseline tells the agent \emph{what is wrong} but not \emph{what to write instead}, in either channel.

\subsection{Results}\label{results}

We ran 10 adversarial tasks $\times$ 3 error-detail modes $\times$ 3 runs $\times$ 3 LLMs (claude-haiku-4-5, claude-sonnet-4-6, gpt-4o-mini) on the enhanced task library, 30 attempts per (model, mode) cell. Every task needs context-dependent validation LLMs cannot satisfy from training-data priors alone, including certified brands, cuisine rules, technique constraints, and exact non-standard scaling ratios. All results are post-audit (Section~\ref{leakage-audit}).

\subsubsection{Success Rate Analysis}\label{success-rate-analysis}

Table~\ref{tab:success3x3} reports success rate per (model, mode) cell, $N=30$. Reflective wins on every model, with the headline rate near or above 86\% on Anthropic while gpt-4o-mini reaches only 63.3\%. The framework lifts the floor for cheaper models but does not erase the capability gap. The Verbose-vs-Reflective gap is wide on Anthropic ($+36.7$ to $+40.0$pp) and narrow on gpt-4o-mini ($+13.3$pp), foreshadowing the per-model significance split (Table~\ref{tab:fisher}). The Verbose-vs-Traditional gap is large on every model ($+30$ to $+50$pp). Even prose-only diagnoses beat generic errors decisively, regardless of model.

\begin{table*}[t]
\centering
\small
\caption{Success rate (\%) by model and error-detail mode. $N=30$ per cell. ``$\Delta$'' columns report percentage-point improvement of reflective over the indicated baseline. Reflective column tinted to mark the highest-success cell per row.}
\label{tab:success3x3}
\renewcommand{\arraystretch}{1.15}
\begin{tabular}{l S[table-format=2.1] S[table-format=2.1] >{\columncolor{winnerviolet}}S[table-format=2.1] S[table-format=+2.1, retain-explicit-plus] S[table-format=+2.1, retain-explicit-plus]}
\toprule
{Model} & {Trad.} & {Verbose} & {Reflective} & {$\Delta$ vs Trad.} & {$\Delta$ vs Verbose} \\
\midrule
\texttt{claude-haiku-4-5}  & 10.0 & 60.0 & 96.7 & +86.7 & +36.7 \\
\texttt{claude-sonnet-4-6} & 16.7 & 46.7 & 86.7 & +70.0 & +40.0 \\
\texttt{gpt-4o-mini}       & 20.0 & 50.0 & 63.3 & +43.3 & +13.3 \\
\bottomrule
\end{tabular}
\end{table*}

Verbose-vs-Traditional measures the value of \emph{any} per-rule error message. Reflective-vs-Verbose measures the marginal value of \emph{machine-readable} structured suggestions. Both are large on the Anthropic models. On gpt-4o-mini, Verbose-vs-Traditional is large ($+30.0$pp, $p=0.029$) but Reflective-vs-Verbose ($+13.3$pp) is within sampling noise at $N=30$ (Fisher's exact $p=0.435$, Section~\ref{sec:eval-stats}). We cannot claim structured suggestions add value over plain-English errors on gpt-4o-mini at this sample size.

\subsubsection{Token Efficiency Analysis}\label{token-efficiency-analysis}\label{sec:eval-tokens}

Reflective is the most token-efficient per success on all three models, despite the heavier per-response payload, because it converges in fewer retries (Table~\ref{tab:tokens3x3}). Focus on the Verbose tokens-per-success and Reflective tokens-per-success columns. On Anthropic, Reflective costs roughly half (haiku 2{,}049 vs 3{,}597, sonnet 2{,}504 vs 5{,}387). On gpt-4o-mini they tie (3{,}665 vs 3{,}548). Traditional's column looks deceptively expensive (9k--26k tokens-per-success) but is divided by 3--6 successes per cell, the noisiest estimate in the table and the reason we do not draw conclusions from absolute Traditional values.

\begin{table*}[t]
\centering
\small
\caption{Tokens per successful task by model and mode, with the Verbose-to-Reflective efficiency factor and a 95\% bootstrap CI (10{,}000 resamples over per-attempt rows). ``Trad.'' is divided by very few successes (3--6 of 30) and should be read as illustrative rather than precise. Tinted cell marks the lowest-token (most efficient) mode per row.}
\label{tab:tokens3x3}
\renewcommand{\arraystretch}{1.15}
\begin{tabular}{l S[table-format=5.0] S[table-format=5.0] S[table-format=5.0] l}
\toprule
{Model} & {Traditional} & {Verbose} & {Reflective} & {V/R efficiency [95\% CI]} \\
\midrule
\texttt{claude-haiku-4-5}  & 26603 & 3597 & \cellcolor{winnerviolet} 2049 & $1.76\times$ [1.05, 3.09] \\
\texttt{claude-sonnet-4-6} & 19237 & 5387 & \cellcolor{winnerviolet} 2504 & $2.15\times$ [1.26, 3.99] \\
\texttt{gpt-4o-mini}       &  9471 & \cellcolor{winnerviolet} 3548 & 3665 & $0.97\times$ [0.47, 2.11] \\
\bottomrule
\end{tabular}
\end{table*}

On both Anthropic models the efficiency factor's 95\% CI sits above $1.0\times$ ($1.76\times$ on haiku, $2.15\times$ on sonnet), so the per-success saving is not point-estimate noise. On gpt-4o-mini the point estimate is $1.0\times$ ($0.97\times$) and the CI straddles unity ([0.47, 2.11]). A real-but-small effect and no effect look the same here, matching the success-rate finding. The Traditional column divides by 3--6 successes per cell and is the noisiest estimate. We report it for completeness but do not draw conclusions from absolute Traditional values.

\subsubsection{Best Demonstration: Triple Challenge (adv\_combo\_001)}\label{best-demonstration-triple-challenge-adv_combo_001}

The qualitative gem is \texttt{adv\_combo\_001}. Traditional and Verbose both score 0/9 across all three models, Reflective scores 7/9. This is the cleanest demonstration in the suite of the marginal value of machine-readable suggestions over plain-English diagnoses. The task fires three context-dependent validation failures at once.

\emph{Task.} Convert recipe to French cuisine + celiac-safe + fix vague measurements.

\emph{Validation errors triggered.}
\begin{itemize}
\item \texttt{INCOMPATIBLE\_INGREDIENT}: coconut milk is not traditional in French cuisine.
\item \texttt{UNSAFE\_FOR\_CELIAC}: generic ``gluten-free flour'' is not certified for celiac safety.
\item \texttt{VAGUE\_MEASUREMENT}: ingredients with terms like ``a handful of'' cannot be scaled.
\end{itemize}

\emph{Per-model outcomes} (3 runs each).
\begin{itemize}
\item \texttt{claude-haiku-4-5}: traditional 0/3, verbose 0/3, reflective 3/3.
\item \texttt{claude-sonnet-4-6}: traditional 0/3, verbose 0/3, reflective 3/3.
\item \texttt{gpt-4o-mini}: traditional 0/3, verbose 0/3, reflective 1/3.
\end{itemize}

Pooled: 0/9 traditional, 0/9 verbose, 7/9 reflective. The reflective payload carries a separate suggestion per error, each naming the specific replacement (e.g., \texttt{with\_specific\_brand: "Bob's Red Mill Certified Gluten Free 1-to-1 Baking Flour"}). The verbose payload carries the same three diagnoses but no brand names, no replacement strings, no precomputed amounts. The agent has to guess each fix from prose, and on this multi-error task none of the three models can.

\subsubsection{Retry Efficiency}\label{retry-efficiency}

Mean retry counts drop monotonically as feedback structure grows. Within the 5-retry budget, pooled per mode across all three models, Traditional averages 4.0--4.6 retries (most exhaust the budget), Verbose 2.6--2.8, Reflective 1.3--2.0. The success-rate gains are not bought by ``trying more''---Reflective succeeds more often \emph{and} retries less often than either baseline.

\subsubsection{Statistical assessment}\label{sec:eval-stats}

With $N=30$ per cell and an unpaired binary outcome, we use Fisher's exact (two-sided) for each pairwise comparison. The OR and $p$-value structure in Table~\ref{tab:fisher} carries the paper's most consequential split. In the Refl.-vs-Verbose columns, both Anthropic models clear $p<0.005$ with odds ratios above 7, while gpt-4o-mini sits at $p=0.435$ with OR$=1.7$. The other two comparisons (Refl.-vs-Trad., Verbose-vs-Trad.) are significant on every model. Read together, prose diagnoses universally beat generic errors, but \emph{structure} on top of prose only pulls clear of noise on capable models at $N=30$.

\begin{table*}[t]
\centering
\small
\caption{Fisher's exact two-sided $p$-values for pairwise success-rate comparisons within each model. ``OR'' is the odds ratio of the first vs the second mode. Tinted $p$-values are significant at $\alpha=0.05$.}
\label{tab:fisher}
\renewcommand{\arraystretch}{1.15}
\begin{tabular}{l rl rl rl}
\toprule
 & \multicolumn{2}{c}{Refl.\ vs Trad.} & \multicolumn{2}{c}{Refl.\ vs Verbose} & \multicolumn{2}{c}{Verbose vs Trad.} \\
\cmidrule(lr){2-3} \cmidrule(lr){4-5} \cmidrule(lr){6-7}
Model & {OR} & {$p$} & {OR} & {$p$} & {OR} & {$p$} \\
\midrule
\texttt{haiku-4-5}   & 261.0 & \cellcolor{winnerviolet}$2.4{\times}10^{-12}$ & 19.3 & \cellcolor{winnerviolet}$0.0011$  & 13.5 & \cellcolor{winnerviolet}$9.4{\times}10^{-5}$ \\
\texttt{sonnet-4-6}  &  32.5 & \cellcolor{winnerviolet}$7.0{\times}10^{-8}$  &  7.4 & \cellcolor{winnerviolet}$0.0022$  &  4.4 & \cellcolor{winnerviolet}$0.025$ \\
\texttt{gpt-4o-mini} &   6.9 & \cellcolor{winnerviolet}$0.0014$              &  1.7 & $0.435$                          &  4.0 & \cellcolor{winnerviolet}$0.029$ \\
\bottomrule
\end{tabular}
\end{table*}

\emph{Interpretation.} Reflective beats Traditional on every model ($p \le 0.0014$). Reflective beats Verbose on both Anthropic models ($p=0.0011$ on haiku, $p=0.0022$ on sonnet) but \emph{not} on gpt-4o-mini ($p=0.435$). The $+13.3$pp gap on gpt-4o-mini is within sampling noise at $N=30$. Verbose beats Traditional on every model ($p \le 0.029$). The contribution unique to \emph{structure}, over plain-English errors, is decisively positive on Anthropic and unproven on gpt-4o-mini. Section~\ref{external-validity} discusses why (model capability, scale, and the per-task pattern in which gpt-4o-mini sometimes recovers a plausible fix from prose where the Anthropic models do not).

\subsection{Context-Dependent Validation Effectiveness}\label{context-dependent-validation-effectiveness}

The experiment only tests validation that needs domain-specific knowledge (Section 1.1). The results bear out the framing. Effectiveness varies measurably across validation types.

\subsubsection{Validation Type Performance Analysis}\label{validation-type-performance-analysis}

\begin{figure}[h]
\centering
\includegraphics{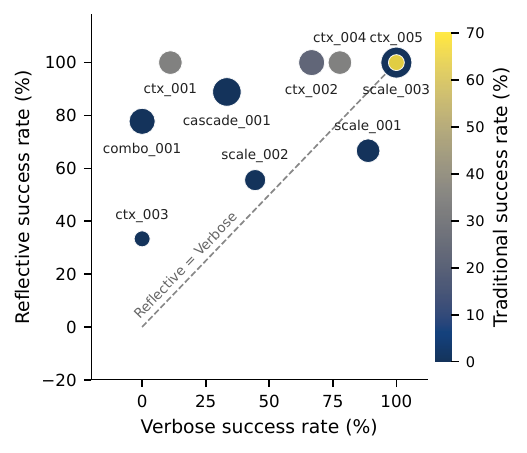}
\caption{Per-task success rate, Verbose vs Reflective, pooled across three models ($N=9$ per task per mode). Dot color = Traditional success rate (darker = higher), dot size = Reflective minus Traditional gap. Tasks above the dashed identity line are Reflective wins. \texttt{scale\_001} (below the line) is the one task where Reflective trails Verbose. Tasks where Traditional and Verbose both score 0 and only Reflective recovers (\texttt{combo\_001}, \texttt{cascade\_001}) appear as large pale dots high on the y-axis.}
\label{fig:per-task}
\end{figure}

Figure~\ref{fig:per-task} plots per-task Verbose vs Reflective success rate, pooled across the three models (9 attempts per task per mode). The full per-task table ships with the repo (\texttt{APPENDIX\_A\_task\_results.md}). Three regions stand out. Tasks where Reflective wins decisively over both baselines (large dots in the upper-left, e.g., \texttt{combo\_001}, \texttt{cascade\_001}, \texttt{ctx\_001}). Tasks where Verbose already saturates and Reflective adds nothing (upper-right corner, e.g., \texttt{ctx\_005}). And \texttt{scale\_001}, the one task where Reflective scores below Verbose (below the identity line). The action-verb-bottlenecked \texttt{ctx\_003} sits as a small dark dot in the lower-left where all three modes flat-line.

\noindent Per validation type.
\begin{itemize}
\tightlist
\item \texttt{INCOMPATIBLE\_INGREDIENT} (cuisine and recipe-style rules, 4 tasks). Reflective wins decisively on \texttt{adv\_ctx\_001/002/004} and \texttt{adv\_combo\_001}. It saturates with Verbose on \texttt{adv\_ctx\_005} (meringue) where the concept name in the diagnosis is itself a complete instruction. Small but non-zero on \texttt{adv\_ctx\_003} (3/9 Refl vs 0/9 each baseline) where action-verb ambiguity bottlenecks all modes.
\item \texttt{UNSAFE\_FOR\_CELIAC} (\texttt{adv\_ctx\_001} and \texttt{adv\_cascade\_001}). The cleanest single Verbose-vs-Reflective gaps. Certified-brand requirements are precisely the case where the diagnosis cannot name the fix (Verbose 1/9 and 3/9 vs Reflective 9/9 and 8/9).
\item \texttt{SCALING\_PRECISION} (3 tasks). Verbose 4--9/9, Reflective 5--9/9, Traditional 0/9. On \texttt{adv\_scale\_001} Verbose (8/9) actually exceeds Reflective (6/9). The diagnosis gives capable models enough to recompute the targets without the structured payload.
\end{itemize}

\emph{Why Traditional sometimes succeeds.} The tasks where Traditional scores non-zero (\texttt{adv\_ctx\_001/002/004/005}) sit at the boundary between generic and context-dependent knowledge. A sufficiently capable model occasionally guesses a plausible fix from a generic error (volunteering ``Bob's Red Mill'' for celiac, or ``starter culture'' for sourdough) when the recipe context cues a familiar pattern. That is why Traditional is the floor and Verbose-vs-Reflective is the core test. Generic errors leak just enough through retry trial-and-error that some tasks recover by chance, but the recovery is unreliable across models and tasks.

\subsubsection{Interpreting the Core Results}\label{interpreting-the-core-results}

The results partially support the hypothesis that structured feedback drives more efficient agent interactions when validation requires context-dependent knowledge, under a stricter comparison than prior single-mode pilots. After the audit stripped all literal fix recipes from \texttt{validation\_errors[].message}, plain-English diagnoses (Verbose) carry only the symptom. Only Reflective's \texttt{recovery\_feedback.suggestions[]} carries the parameters to act on. Under that separation, Reflective beats Verbose by $+36.7$pp on haiku ($p=0.0011$) and $+40.0$pp on sonnet ($p=0.0022$). The $+13.3$pp gap on gpt-4o-mini is not significant ($p=0.435$). Structured suggestions over plain-English errors are robustly positive on Anthropic, unproven on gpt-4o-mini at $N=30$.

Token efficiency tracks the same split. On Anthropic, per-success cost drops sharply from Verbose to Reflective (haiku $1.8\times$, sonnet $2.2\times$). Machine-readable suggestions are cheaper than blind retries when they actually shorten the retry chain. On gpt-4o-mini, Verbose and Reflective per-success token costs tie ($3{,}548$ vs $3{,}665$), matching the smaller success-rate gap. The $43$--$87$pp lift from Traditional to either Verbose or Reflective is unambiguous on every model. \emph{Some} per-rule diagnostic beats a generic error string, regardless of model.

\subsubsection{Fairness of Comparison}\label{fairness-of-comparison}

All three modes face identical validation logic and identical task inputs. The only differences are (a) verbosity of the per-error \texttt{message} field and (b) whether the response carries a \texttt{recovery\_feedback.suggestions[]} payload with machine-readable parameters. After the leakage audit (Section~\ref{leakage-audit}), no literal fix recipe (target brand, exact amount, replacement string) appears in the Verbose payload, only inside Reflective's \texttt{recovery\_feedback}. Verbose-vs-Reflective is therefore a clean test of whether structure matters \emph{given} the same diagnostic information in prose.

Traditional's low success rate (10--20\%) reflects blind trial-and-error in a specialized domain. Verbose shows what plain-English diagnoses alone recover (46.7--60\%). Reflective shows the additional lift from machine-readable suggestions (63.3--96.7\%). The Verbose-to-Reflective gap is the framework's intrinsic contribution beyond mere verbosity, large and significant on Anthropic, small and non-significant on gpt-4o-mini.

\subsection{Limitations and Threats to Validity}\label{limitations-and-threats-to-validity}

\subsubsection{Internal Validity}\label{internal-validity}

Token counts come from provider APIs (Anthropic Messages \texttt{usage}, OpenAI \texttt{usage} via LangChain), so the per-success ratios in Table~\ref{tab:tokens3x3} reflect billed tokens, not character estimates. The largest residual internal-validity threat is the audit itself. We removed every literal fix string we could find from \texttt{validation\_errors[].message}, but the diagnostic phrasing (``coconut milk is not traditional in French cuisine'') still narrows the search space for a competent LLM. We treat this as a feature of the Verbose baseline, not a bug (it is what an honest plain-English error message looks like), which makes the Verbose-vs-Reflective gap conservative against Reflective.

\paragraph{Researcher degree of freedom: how verbose is ``Verbose''?} The exact wording of the Verbose diagnoses is an experimenter choice. A terser Verbose (just the error code, \texttt{INCOMPATIBLE\_INGREDIENT}) would widen the Verbose-vs-Reflective gap by stripping the natural-language hint that primes capable models. A richer Verbose (``a traditional French dairy substitute is required'') would narrow it. We chose phrasings a thoughtful production API author would write \emph{after} being told not to leak the literal answer. Specific enough to communicate the failure, vague enough not to dictate the fix. Other choices would shift the headline numbers. We disclose this rather than claim our wording is uniquely correct, and we ship the exact diagnosis strings in the validator source for reproducibility.

\subsubsection{External Validity}\label{external-validity}

Several factors limit generalizability.

\paragraph{Single domain} All 10 tasks come from one validator (recipe conversion). The Verbose-to-Reflective benefit likely depends on how cleanly the domain decomposes into discrete fix actions (REPLACE\_INCOMPATIBLE\_INGREDIENT, USE\_SPECIFIC\_BRAND, CLARIFY\_MEASUREMENT, MODIFY\_PARAMS). Domains with less crisp action vocabularies (open-ended rewriting, subjective grading, free-form planning) may show smaller gaps.

\paragraph{Adversarial task selection} Tasks fail strict validation in ways that require API-side knowledge. Real traffic is mostly valid requests, where the Verbose-vs-Reflective gap collapses to zero (no errors imply no recovery payload). Read the deltas as per-failure recovery lift, not average per-call lift on a production traffic mix.

\paragraph{Single agent architecture} A simple retry-loop agent capped at 5 attempts. More capable agents (ReAct with tool-use planning, multi-agent critics, agents with memory across episodes) might reason their way from the Verbose diagnosis to the literal fix the Reflective payload hands over directly. We expect this narrows the gap on capable Anthropic models more than on gpt-4o-mini.

\paragraph{Three-model vendor coverage} Three LLMs from two vendors rules out single-model artifacts but cannot characterize how the gap varies with model capability in general. Reflective-vs-Verbose figures: $+36.7$pp on haiku-4-5 ($p=0.0011$), $+40.0$pp on sonnet-4-6 ($p=0.0022$), $+13.3$pp on gpt-4o-mini ($p=0.435$, not significant). Three explanations fit the gpt-4o-mini null and we cannot tell them apart at this scale. (a) It extracts more from prose than the Anthropic models. (b) It ignores the structured payload and reasons from prose either way. (c) The true gap exists but is too small to detect at $N=30$. Whether GPT-4-class models close, restore, or invert the gap is open.

\paragraph{Scale} 30 attempts per (model, mode) cell is enough for Fisher's exact (Section~\ref{sec:eval-stats}) but small for per-task or per-error-type slicing. We avoid strong conclusions from any single-task cell.

\subsubsection{Domain Portability: Replication on a Second API}\label{domain-portability}

\paragraph{The recipe domain is the credibility anchor} The gpt-4o-mini Verbose-vs-Reflective gap is honestly non-significant ($p=0.435$, $+13.3$pp). The recipe pilot reports a null on one of three models for the core comparison, and the null is load-bearing. It shows the experiment is sensitive enough to \emph{not} find an effect when the effect is too small to detect at $N=30$, and it sets an honest prior for what a second domain should show. Treat replications that produce only positive results with skepticism. The recipe results earn trust precisely because one cell did not cooperate.

To partially address the single-domain threat, we replicated the full 3-arm $\times$ 3-model protocol on a structurally different API. \emph{Acme billing} is a fictitious refund/dispute API on Stripe's sandbox. Where recipe involves free-text ingredient lists, fuzzy cuisine rules, and continuous numeric scaling, Acme involves transactional state, hard monetary constraints (refund caps, approval-token requirements), and a discrete policy ruleset. The same audit script (\texttt{experiments/audit\_prompt\_leakage.py}, Section~\ref{leakage-audit}) ran on Acme's validator messages and task prompts. Results below are post-audit. Both APIs and audits are open-source.

\begin{figure*}[t]
\centering
\includegraphics[width=\textwidth]{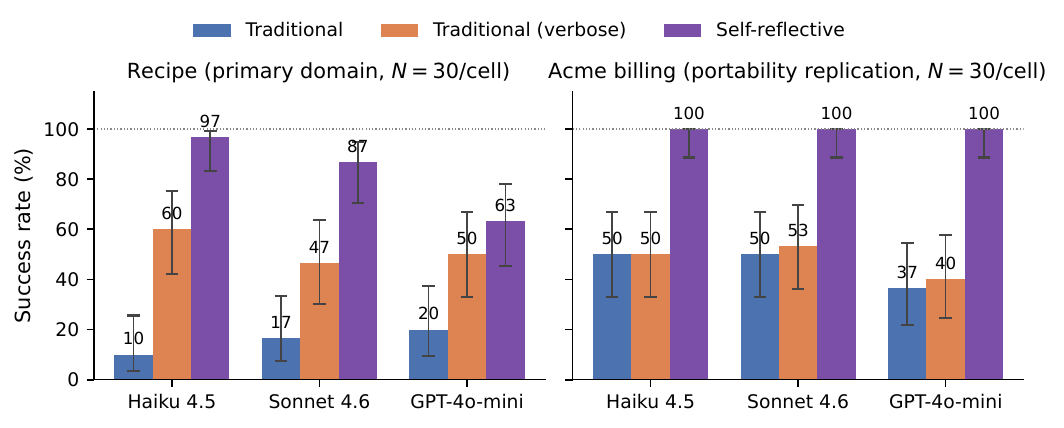}
\caption{Domain portability: same 3-arm protocol on two structurally different APIs. Recipe (left) is the credibility-anchor result. The gpt-4o-mini Verbose-vs-Reflective gap is honestly non-significant ($p=0.435$), showing the experiment's sensitivity to null results. Acme billing (right) reproduces the Reflective $\gg$ \{Traditional, Verbose\} pattern with reflective at the 100\% ceiling on every model. Read the uniform 100\% as a portability check, not as a stronger effect-size estimate. The author-constructed-rules caveat in the text applies.}
\label{fig:portability}
\end{figure*}

\paragraph{Headline (Acme, $N=30$ per cell, 270 attempts total).} Traditional 45.6\% [35.7, 55.8], Verbose 47.8\% [37.8, 58.0], Reflective 100.0\% [95.9, 100.0] (Wilson 95\% CIs, combined across models). Fisher's exact gives Traditional vs Reflective $p=3.75\mathrm{e}{-19}$, Verbose vs Reflective $p=3.64\mathrm{e}{-18}$, Traditional vs Verbose $p=0.881$ (no benefit from prose alone). Reflective hits 30/30 on every model. The Verbose-vs-Reflective gap is $+50.0$pp on haiku, $+46.7$pp on sonnet, $+60.0$pp on gpt-4o-mini. Figure~\ref{fig:portability} shows the two domains side by side. gpt-4o-mini, with no significant lift on recipe ($p=0.435$), reaches 100\% reflective on Acme. That fits the interpretation that the recipe null was a power/scale issue, not a model-level ceiling. We cannot rule out that Acme's rules decompose more cleanly into discrete fix actions.

\paragraph{Caveat: the 100\% ceiling and author-constructed rules} Acme's rules are hand-authored by us. A 100\% cell with $p<10^{-18}$ is exactly what a skeptical reader should expect when rule-authors and method-authors are the same people. The structured suggestion field encodes information very close to the answer key. Two safeguards apply. (a) The leak audit catches literal fix-value substrings in validator messages and prompts. (b) The Verbose arm receives the same per-rule diagnoses minus only the machine-readable suggestion object. Neither rules out that we unconsciously picked Acme rules that break down unusually cleanly. Recipe is the more credible result \emph{because} it contains a null. Read Acme's uniform 100\% as a portability check, not a stronger effect-size estimate.

\paragraph{What this replication does and does not show} The structured-suggestions $\gg$ generic and structured-suggestions $\geq$ verbose pattern reproduces on an API with very different surface semantics, using the same agent runner, models, and audit methodology. It does not show generalization to APIs the authors did not build. The strongest external-validity evidence would come from a third-party API with native context-dependent validation (Stripe's own dispute reasoning, GitHub's branch-protection errors, Kubernetes admission-webhook denials) retrofitted with a Reflective response field. Section~\ref{future-research-directions} flags this as the most important open evaluation.

\subsubsection{Construct and Conclusion Validity}\label{construct-validity}\label{conclusion-validity}

Success rate, tokens-per-success, and retry count directly measure the constructs. We do not measure latency, feedback-quality characteristics (actionability, completeness), or cross-episode learning. Statistical claims stay at the (model, mode) cell level. Per-task slices have at most 9 attempts pooled across models, so we avoid per-task significance tests.

\section{Discussion and Future Work}\label{discussion-and-future-work}

Self-reflective APIs deliver measurable, production-relevant benefits in specialized domains with context-dependent validation, under a stricter comparison than prior single-mode pilots. After the audit stripped literal fix recipes from plain-English error messages (Section~\ref{leakage-audit}), structured \texttt{recovery\_feedback.suggestions[]} still beats verbose prose by $+36.7$pp on haiku-4-5 ($p=0.0011$) and $+40.0$pp on sonnet-4-6 ($p=0.0022$), with 1.8--2.2$\times$ better per-success token efficiency. On gpt-4o-mini the gap ($+13.3$pp, $p=0.435$) is not significant at $N=30$ and per-success token cost ties. Below we cover when the lift lands, when it does not, and what is open.

\subsection{When Self-Reflective Feedback Provides Value}\label{when-self-reflective-feedback-provides-value}

The headline lesson is that reflective feedback pays off precisely when APIs hold domain-specific knowledge LLMs lack. This sharpens where to spend on self-reflective patterns.

Per-task patterns (pooled across three models, 9 attempts per task per mode) sharpen the picture.

\paragraph{Cuisine authenticity} On \texttt{adv\_ctx\_002} (French/coconut milk), the diagnosis ``coconut milk is not traditional in French cuisine'' is enough for capable models to often guess a dairy replacement, and the explicit \texttt{REPLACE\_INCOMPATIBLE\_INGREDIENT} payload makes it deterministic. Reflective 9/9, Verbose 6/9, Traditional 2/9.

\paragraph{Certified-brand requirements} ``A specific certified gluten-free flour brand is required'' tells the agent something is wrong but not which brand to write. On \texttt{adv\_ctx\_001} (celiac flour), \texttt{USE\_SPECIFIC\_BRAND} with \texttt{Bob's Red Mill Certified GF} closes that gap, producing Reflective 9/9 against Verbose 1/9 and Traditional 3/9, one of the largest Verbose-vs-Reflective gaps in the suite.

\paragraph{Cascading multi-step validation} Each fix exposes the next error, and structured suggestions accelerate convergence within the 5-retry budget. \texttt{adv\_cascade\_001} (fix flour brand $\to$ then oats brand) lands at Reflective 8/9, Verbose 3/9, Traditional 0/9.

\paragraph{Triple challenge} Reflective 7/9, Verbose 0/9, Traditional 0/9 on \texttt{adv\_combo\_001} (cuisine + celiac + vague measurements). With three independent error types firing simultaneously, only the structured payload guides the agent through all three fixes, and both baseline modes are blanked.

\paragraph{Recipe-style technique constraints} When the diagnosis itself names a well-known concept (``meringue should not contain a chemical leavener like baking powder''), Verbose already saturates and Reflective adds nothing. \texttt{adv\_ctx\_005} ties at Reflective 9/9 and Verbose 9/9, against Traditional 6/9.

\paragraph{Non-standard scaling ratios} Once the validator exposes the per-ingredient \emph{provided\_amount} in the diagnosis, capable models often recompute the target without needing the explicit \texttt{expected\_amount} from the structured payload. Across \texttt{adv\_scale\_001/002/003} Verbose lands 8/9, 4/9, 9/9 and Reflective 6/9, 5/9, 9/9 (Traditional 0/9 each), and on \texttt{adv\_scale\_001} Verbose actually edges out Reflective (8/9 vs 6/9).

\paragraph{Where to invest} Build self-reflective feedback for proprietary business logic LLMs cannot know (discontinued products, current inventory, company policies) and for real-time state dependencies (user profiles, quotas, session data). It also pays for specialized domain requirements (medical certifications, regulatory compliance, cultural standards), multi-step cascading validation where fixing A reveals B, and precise computational requirements paired with capable models.

\paragraph{Where to skip it} Skip self-reflective feedback for generic validation LLMs already handle (``add gluten-free flour for gluten-free recipe''), straightforward successful requests, and low-stakes domains where trial-and-error is acceptable.

\subsection{When Self-Reflection Fails or Saturates}\label{when-self-reflection-fails}

Three cell types in the pilot expose where Reflective does not deliver its full advantage. Each has a different design implication.

\subsubsection{Verbose saturates (no headroom left)}\label{verbose-saturates}

When the diagnosis names a familiar concept the model can act on, Verbose ties Reflective and the structured payload is redundant. On \texttt{adv\_ctx\_005} (meringue + baking powder), Reflective scores 9/9, Verbose 9/9, Traditional 6/9. The prose ``a meringue should not contain a chemical leavener like baking powder'' is already a complete instruction for any model that knows what a meringue is. The engineering effort belongs on rules whose names do not name the fix, such as certified brands, exact non-standard amounts, and replacement-by-cuisine rules.

\subsubsection{Model can recompute the answer from a diagnosis (Verbose ties Reflective)}\label{verbose-ties-reflective-via-recompute}

When the diagnosis exposes enough machine-checkable substrate that the model can do the arithmetic itself, Verbose reaches Reflective. The \texttt{adv\_scale\_*} family makes this visible. Verbose 4--9/9, Reflective 5--9/9, Traditional 0/9 across three scaling tasks, with Verbose (8/9) edging Reflective (6/9) on \texttt{adv\_scale\_001}. \texttt{SCALING\_PRECISION\_REQUIRED} lists ``ingredient X has provided\_amount A but ratio R requires A$\times$R'' without writing the product. Capable models multiply, and the structured \texttt{expected\_amount} becomes redundant. The gap depends on how much arithmetic or string composition the diagnosis offloads. More raw substrate means a smaller Reflective lift. Pick the line per error type, trading honest diagnostics against guaranteed recovery.

\subsubsection{Structured guidance occasionally hurts (Reflective $<$ Verbose)}\label{reflective-below-verbose}

On a small number of tasks, Reflective scores \emph{lower} than Verbose, because the structured suggestion misleads the agent or competes with a correct interpretation the agent would have reached from prose alone. The clearest case is \texttt{adv\_scale\_001} (Verbose 8/9, Reflective 6/9). The \texttt{FIX\_SCALING\_PRECISION} payload exposes per-ingredient \texttt{expected\_amount} fields, and we see agent runs mis-merge these, applying the new amount to a different ingredient than the suggestion targeted, or treating a single-ingredient suggestion as a full ingredient-list rewrite. Given just the prose diagnosis, the same agent recomputes the targets and writes them in the original schema correctly.

Across 10 tasks pooled across three models (per-task table in \texttt{APPENDIX\_A\_task\_results.md}), Reflective $\geq$ Verbose on 9 tasks and $<$ Verbose on 1 (\texttt{adv\_scale\_001}, $-2/9$). The aggregate lift stays positive, but structured suggestions are not strictly dominant per task. Structured payloads carry a \emph{schema interpretation cost}. When the diagnosis already names a familiar concept (Section~\ref{verbose-saturates}) or already exposes the substrate to recompute (Section~\ref{verbose-ties-reflective-via-recompute}), structure can introduce parse errors prose did not have. Do not assume Reflective dominates Verbose monotonically. A/B per validator before deploying universally.

\subsubsection{The action-verb bottleneck: parameter values without operational semantics}\label{reflective-not-ceiling}

Even with a structured \texttt{REPLACE\_INCOMPATIBLE\_INGREDIENT} payload, agents sometimes add the suggested replacement instead of substituting it, producing a duplicate-ingredient failure on the next round. \texttt{adv\_ctx\_003} (traditional Italian + vegan cheese) is the canonical example, with Reflective 3/9, Verbose 3/9, Traditional 3/9, flat across modes. The structured payload says ``replace vegan cheese with nutritional yeast,'' but the agent often emits both ingredients, or omits vegan cheese without adding the replacement.

The payload provides \emph{parameter values} (what to put in the field), not \emph{operational semantics} (how to transform current state into target state). \texttt{REPLACE\_INCOMPATIBLE\_INGREDIENT} names the operation, but interpreting it requires the agent to model the ingredient list as a set (remove one element, insert another atomically) rather than treat the suggestion as an additive patch. The agent is not missing information about \emph{what} to write. It is failing to reason correctly about \emph{how} to write it. This is the \emph{grounded action execution} problem from embodied AI and robotics, where systems receive correct high-level instructions but fail to translate them into correct low-level state transitions \cite{ref10}. In the API setting, the LLM must map a suggestion schema entry to a JSON edit on the request body. The current schema specifies what to submit, not how to derive it from the previous request, leaving the mapping fully on the model. Capable instruction-followers (claude-haiku, claude-sonnet) handle it reliably. gpt-4o-mini and \texttt{adv\_ctx\_003} expose the failure mode.

Two complementary fixes follow. \emph{Transactional diff schemas} specify a patch over previous request state (``in \texttt{ingredients[]}, find element matching \texttt{vegan\_cheese} and replace with \texttt{nutritional\_yeast}'', like git diff or JSON Patch RFC 6902), so the model no longer tracks state. \emph{Before/after state} includes both expected current and desired target states so the agent can verify it is patching the right field. The tradeoff is payload size and validator coupling. Payload size is already modest ($\sim$11\% overhead), so coupling is the bigger constraint. The design implication is to distinguish two classes of suggestion by where the bottleneck lies. \emph{Parameter-value suggestions} (knows what operation, not what value, like certified brands, exact scaling targets, replacement-by-cuisine) and \emph{operational-semantic suggestions} (knows what value, not how to apply it). The current schema serves the first class well. The second needs diff-style schemas or explicit before/after state. A flat \texttt{adv\_ctx\_003}-style result across all modes is the signal that a task has crossed into the second class. A/B value-style vs diff-style per validator before deploying.

\subsubsection{Model-capability ceiling}\label{model-capability-ceiling}

gpt-4o-mini's Reflective rate (63.3\%) sits well below the Anthropic models' (96.7\% and 86.7\%). The ceiling is not the framework (the same payloads recover near-all attempts on Anthropic) but the smaller model's tool-use and instruction-following reliability. Self-reflective APIs raise the floor for cheaper models but do not erase the gap to capable ones, and the Verbose-vs-Reflective marginal lift on gpt-4o-mini is not distinguishable from zero at $N=30$.

\subsubsection{Genuinely impossible tasks (out of scope for this pilot)}\label{impossible-tasks-out-of-scope}

Pilot tasks fail validation \emph{recoverably}. Production traffic also has genuinely impossible requests (chemical impossibility, contradictory constraints, resource unavailability). The framework should communicate impossibility via a \texttt{NO\_RECOVERY\_AVAILABLE} suggestion type rather than emit plausible but unfollowable fixes. We do not measure this here.

\subsubsection{Token Efficiency Pattern}\label{the-token-efficiency-paradox}

Per-attempt token cost is roughly flat across modes. Per-success cost diverges because Traditional attempts mostly burn the 5-retry budget without converging. Verbose recovers in 1--2 retries when the diagnosis names a known concept. Reflective recovers the same cases plus the long tail (specific brands, exact targets, cuisine-specific replacements) at comparable per-attempt size, so per-success cost drops (Table~\ref{tab:tokens3x3}).

\subsection{Future Research Directions}\label{future-research-directions}

The most important open question is a \emph{retrofit on a third-party API the authors did not build}, such as Stripe dispute reasoning, GitHub branch-protection denials, or Kubernetes admission webhooks, to test whether the Reflective $\gg$ Verbose pattern survives outside author-constructed validators (Section~\ref{domain-portability}). Two further directions follow. \emph{Capability-tier characterization} would map how the gap varies with model capability, since the gpt-4o-mini null leaves open whether smaller models extract less from structured payloads or whether the gap is a power-at-$N{=}30$ artifact. \emph{Transactional diff schemas} (Section~\ref{reflective-not-ceiling}) would test whether suggestion payloads expressed as JSON Patch-style state transitions close the action-verb bottleneck on tasks like \texttt{adv\_ctx\_003}. Secondary directions include standardization (OpenAPI extensions, common action vocabularies), multi-round clarification protocols, and security implications of exposing validator internals.

\section{Conclusion}\label{conclusion}

A \emph{self-reflective API} returns, on validation failure, a machine-readable \texttt{recovery\_feedback.suggestions[]} payload sufficient for an autonomous agent to repair the request and retry. On a leak-audited pilot of 10 adversarial recipe-conversion tasks $\times$ 3 error-detail modes $\times$ 3 LLMs $\times$ 3 runs ($N{=}30$ per cell), this payload beats plain-English diagnoses by $+36.7$pp on haiku-4-5 ($p{=}0.0011$) and $+40.0$pp on sonnet-4-6 ($p{=}0.0022$), with $1.8$--$2.2\times$ better per-success token efficiency. On gpt-4o-mini the $+13.3$pp gap is not significant ($p{=}0.435$). We report it honestly as the credibility anchor for the recipe domain. A second-domain replication on an Acme billing API confirmed the pattern. Reflective beat both Traditional and Verbose on all three models, including gpt-4o-mini, which had produced the null on recipe.

Three pieces a downstream paper can cite independently. The \textbf{Self-Reflective API Schema v0.1} (Section~\ref{schema-v0-1}), a named minimal wire-format contract. The three-mode \emph{generic / verbose / reflective} protocol that isolates the value of \emph{structure} from mere verbosity. And the leakage-audit methodology and tooling (\texttt{audit\_prompt\_leakage.py}) for any LLM benchmark whose validator and tasks come from the same team. Reflective wins decisively where the diagnosis cannot name the fix (certified brands, cuisine-specific replacements, cascading multi-step rules), and ties or marginally trails Verbose where the diagnosis itself names a familiar concept or exposes enough substrate for the model to recompute. Structure beats verbosity in domains where APIs hold knowledge advantages over general-purpose LLMs.

\subsection*{Use of AI Assistance}
The experimental runner, analysis scripts, and figure-generation code were developed with Claude Code under the authors' direction and review. The paper was drafted and copy-edited with Claude assistance. All claims, numbers, and figures were verified by the authors against the underlying data.

\begin{acks}
We thank Ananth Grama, Yifan Wang, and Arun Ramamurthy for early discussions that helped sharpen the framing of self-reflective APIs and for feedback on the experimental design.
\end{acks}

\end{document}